\documentstyle[epsf,12pt]{article}
\title{Moving Embedded Solitons}

\author{Alan R. Champneys \vspace{0.1cm}\\
Department of Engineering Mathematics, The University of Bristol,\\
Bristol BS8 1TR, United Kingdom \vspace{0.4cm} \\
Boris A.Malomed \vspace{0.1cm} \\
Department of Interdisciplinary Studies, Faculty of Engineering, Tel Aviv
University, \\
Tel Aviv 69978, Israel}
\date{\today To appear in J Phys A.}
\begin{document}
\maketitle

\begin{abstract}
The first theoretical results are reported predicting {\em moving}
solitons residing inside ({\it embedded} into) the continuous spectrum of
radiation modes. The model taken is a Bragg-grating medium with Kerr
nonlinearity and additional second-derivative (wave) terms. The
moving embedded solitons (ESs) are doubly isolated (of codimension 2),
but, nevertheless, structurally stable. 
Like quiescent ESs, moving ESs are argued
to be stable to linear approximation, and {\it semi}-stable nonlinearly.
Estimates show that moving ESs
may be experimentally observed as $\sim$10 fs pulses with velocity 
$\leq 1/10$th that of light.

\begin{center} 
\end{center}

\end{abstract}

Recent studies have revealed a novel type of soliton (``solitary wave'' is
more accurate since we do not assume integrability) that is {\it embedded}
into the continuous spectrum, i.e., the soliton's internal frequency is in
resonance with linear (radiation) waves. Generally, such a soliton should
not exist, one finding instead a ``quasi-soliton'' with non-vanishing
oscillatory tails (radiation component) \cite{Boyd}. Nevertheless, {\em bona
fide} (exponentially decaying) solitons can exist as {\it codimension-one}
solutions if, at discrete values of the (quasi-)soliton's internal
frequency, the amplitude of the tail exactly vanishes, while the soliton
remains embedded into the continuous spectrum. This requires the spectrum of
the corresponding linearized system to consist of (at least) two branches,
one corresponding to exponentially localized solutions, and the other to
radiation modes. In terms of the travelling-wave ordinary differential
equations (ODEs), the origin must be a {\em saddle-centre} equilibrium.

Examples of such {\it embedded solitons} (ESs) were found in water-wave
models \cite{ChGr} and in several nonlinear-optical ones, e.g., a Bragg
grating with dispersion and/or diffraction terms \cite{we}, and
second-harmonic generation (SHG) in the presence of a self-defocusing Kerr
nonlinearity (\cite{YKM,Trillo}). The term ``ES'' was proposed in Ref. \cite
{YKM}.

It is relevant to stress that ESs, although they are isolated solutions, are 
{\em not} structurally unstable. Indeed, a small change of the model's
parameters will slightly change the location of ES (e.g., its energy and
momentum, see below), but will not destroy it, which is quite obvious from
the already published results \cite{ChGr,YKM}. In this respect, they may be
called generic solutions of codimension one.

ESs are interesting for several reasons, firstly because they frequently
appear when higher-order (singular) perturbations are added to the system,
which may completely change its soliton spectrum (see e.g.\ \cite{we}).
Secondly, optical ESs may have a potential for applications, just because
they are isolated solitons rather than members of continuous families.
Finally, and most crucially for their physical applications, it appears that
ESs are {\em semi}-stable objects. That is, as is proven in Ref.\ \cite{YKM}
analytically in a fairly general form, and checked numerically for a
particular model combining SHG (quadratic) and Kerr (cubic) nonlinearities,
ESs are fully stable in the linear approximation, but are subject to a
slowly growing (sub-exponential) {\em one-sided} nonlinear instability (see
below). The analytical proof of the semistability presented in Ref. \cite
{YKM} applies to any system that gives rise to ESs. As for the one-sided
nonlinear instability, its development depends on values of the system's
parameters; in some cases, it may be developing so slowly that ES, to all
practical purposes, may be regarded as a fully stable object \cite{unpub}.

An issue important both for applications and by itself is whether {\em moving
} ESs (ones with non-zero momentum) may occur in systems where they cannot
be generated by a straightforward transformation, like Galilean or Lorentz
transformation (the absence of the corresponding invariance is typical for
nonlinear-optical systems). The objective of the present work is to search
for moving ESs in a physically important system, viz., a nonlinear
Bragg-grating model similar to that introduced in \cite{we}, which takes
into account second-derivative (wave) terms. In fact (see below), this
system has a broader physical purport than was originally assumed in \cite
{we}. The absence of the Galilean or Lorentzian invariance in it is obvious
because there is a reference frame in which the Bragg grating is quiescent.
Although exact solutions for moving solitons are available in the
traditional version of this model, which neglects the second-derivative
terms \cite{CJAW,gap}, they can be obtained by the Lorentz transformation
from the quiescent solitons only in the limiting case of the Thirring model 
\cite{Thirring}, which is completely integrable \cite{integr}.

We start from a system of partial differential equations (PDEs) governing
evolution of right- ($u(x,t)$) and left- ($v(x,t)$) traveling waves that
continuously transform into each other due to the resonant reflection on the
grating: 
\begin{eqnarray*}
iu_{t}+iu_{x}+(2k)^{-1}\left( u_{xx}-u_{tt}\right) +\left( \sigma
|u|^{2}+|v|^{2}\right) u+v &=&0, \\
iv_{t}-iv_{x}+(2k)^{-1}\left( v_{xx}-v_{tt}\right) +\left( \sigma
|v|^{2}+|u|^{2}\right) v+u &=&0.
\end{eqnarray*}
Here, the cubic and linear cross-coupling terms account, respectively, for
nonlinear cross-phase modulations and Bragg scattering. The most natural
physical value of the relative self-modulation coefficient $\sigma $ is $1/2$
, but it will be quite useful to keep $\sigma $ as an arbitrary positive
parameter. Note that Eqs. (1) and (2) have three natural integrals of
motion: the energy (norm) and momentum, 
\[
E\equiv \int_{-\infty }^{+\infty }\left[ |u(x)|^{2}+|v(x)|^{2}\right]
dx\,,\,\,P\equiv i\int_{-\infty }^{+\infty }\left( u_{x}^{\ast
}u+v_{x}^{\ast }v\right) dx\,,
\]
and a Hamiltonian, an expression for which is obvious.

The energy plays a crucial role in analyzing ES stability \cite{YKM}, as ESs
are isolated solutions with uniquely determined values of the energy. Hence,
any small perturbation which slightly increases the ES's energy is safe,
while a perturbation that slightly decreases the energy triggers a slow
(sub-exponential) decay into radiation. So in this sense, the weak
instability of an ES is one-sided, as mentioned above, and in some cases it
may be {\em extremely} weak \cite{unpub}.

Eqs. (1) and (2) can be derived from the Maxwell's equations for a nonlinear
medium, assuming a superposition of two counter-propagating electromagnetic
waves, $u(x,t)\,\exp (ikx-i\omega t)$ and $v(x,t)\,\exp (-ikx-i\omega t)$,
where the wavenumber $k$ and frequency $\omega $ are related by the linear
dispersion relations (disregarding their Bragg coupling), the functions $
u(x,t)$ and $v(x,t)$ being slowly varying as compared to the carrier waves.
Taking (for simplicity) a medium whose temporal dispersion may be neglected,
and setting $c_{0}\equiv 1$ (hence, $\omega =k$), one derives, to lowest
order in the small parameter $1/2k$, Eqs.\ (1) and (2) without the
second-derivative terms, i.e., a standard model of the Bragg reflector
filled by a Kerr-nonlinear medium \cite{CJAW,gap}. As shown in \cite{we},
the second-derivative (wave) terms which come in at the next order
drastically alter the soliton spectrum of the model (since this is a {\it 
singular perturbation}, increasing the order of the PDEs). In an experiment
(see below for an estimate of physical parameters), the effect of the
additional terms may be seen if the observation time and/or propagation
distance are long enough.

Solitons are solutions of the form 
\[
u(x,t)=\exp (-i\Delta \omega t)\,U(\xi ),\;v(x,t)=\exp (-i\Delta \omega
t)\,V(\xi ),
\]
where $\xi \equiv x-vt$, $v$ is the soliton's velocity, and $\Delta \omega $
is a frequency shift. The substitution of this expression into Eqs. (1) and
(2) yields the ODEs, 
\begin{equation}
\chi U+i(1-c)U^{\prime }+DU^{\prime \prime }+\left( \sigma
|U|^{2}+|V|^{2}\right) U+V=0,  \label{U}
\end{equation}
\begin{equation}
\chi V-i(1+c)V^{\prime }+DV^{\prime \prime }+\left( \sigma
|V|^{2}+|U|^{2}\right) V+U=0,  \label{V}
\end{equation}
where $\chi \equiv \Delta \omega +\left( \Delta \omega \right) ^{2}/2k$, the
effective velocity is $c\equiv (1+\Delta \omega /k)v$, and an effective
dispersion coefficient is $D\equiv \left( 1-v^{2}\right) /2k$.

In \cite{we} the same ODEs were derived in two more special physical
contexts: (i) a nonlinear Bragg-grating medium incorporating {\it 
spatial-dispersion} effects and (ii) spatial evolution (i.e.\ with $t$
realized as a propagation coordinate) in a planar waveguide equipped with a
Bragg grating in the form of a set of parallel scores, taking ordinary
diffraction into regard. While all these systems are described Eqs. (\ref{U}) 
and (\ref{V}), the new physical interpretation of the model as describing
the usual Bragg-grating system with the wave terms taken into regard, seems
most fundamental.

To look for ES solutions, we must first satisfy the necessary condition,
viz., that the linearization of the ODEs should be of the saddle-centre
type. That is, at least one pair of eigenvalues must be purely imaginary
(otherwise, we are dealing with {\em regular}, i.e., non-embedded,
solitons), and at least one pair must {\em not} be purely imaginary
(otherwise, there can be no exponentially decaying tails). Hence the region
in which ESs may exist may be delineated by substituting $U,V\sim \exp
(\lambda \xi )$ into the linearized equations and solving the resulting
eighth-order algebraic equation for $\lambda $ numerically. It is easy to
demonstrate that purely real or imaginary eigenvalues always appear in
pairs, and complex eigenvalues in quadruples: if $\lambda $ is an
eigenvalue, then so are $\pm \lambda $ and $\pm \lambda ^{\ast }$.

We do not display here the full results for the linear spectrum, as they are
rather cumbersome. But note that in the quiescent case ($c=0$) the spectrum
is expressible in a closed form \cite{we}, and the region in the $(\chi ,D)$
-plane where ESs may occur is just $|\chi |<1$, $D>0$. When $c\neq 0$, these
borders to the saddle-centre region of $(c,\chi ,D)$-space retain exactly
the same meaning (but there appear additional bounding surfaces that, in
fact, are {\em not} encountered by any of the ES branches that we have
computed, see below). Two degenerate limits of special interest are $\chi
\rightarrow +1$ (the soliton amplitude going to zero) and $\chi \rightarrow
-1$ (a smooth transition into a regular soliton).

Eqs. (\ref{U}) and (\ref{V}) were numerically solved by means of the same
techniques as used in Ref. \cite{we}. That is, a two-point boundary-value
problem is posed on a long but finite $x$-interval, with boundary conditions
chosen to place the solution in the stable or unstable eigenspaces at the
endpoints \cite{numerical}. The boundary-value problem can be formulated so
that the imaginary parts of $A(\xi )$ and $B(\xi )$ are always even
functions, while the real parts are odd. Using these {\em reversibility}
conditions at the midpoint of the soliton, the numerical problem was posed
more simply on the half $x$-interval. Only {\it fundamental} (single-humped)
solitons were sought because, although multi-humped ESs may easily exist,
they have no chance to be stable \cite{YKM}. Continuation of the solutions
corresponding to variation of relevant parameters was carried out by means
of the well-known software package AUTO \cite{AUTO}.

Quiescent ESs (with $c=0$) in the present model were found in Ref.\ \cite{we}
, aided by the observation that, at $c=0$, Eqs. (\ref{U}) and (\ref{V})
admit an invariant reduction $V\equiv U^{\ast }$, thus reducing the system's
order from $8$ to $4$. The result was that there exist exactly three
different branches of quiescent ES solutions. Because ESs exist at
isolated values of the energy, each branch can be represented by a curve $
E(D)$ in three separate $D$-intervals (which overlap). Equivalently, the
curves can be represented as $D(\chi )$ for $-1<\chi <1$.

To the best of our knowledge, moving ESs have never been found before in any
model. Our numerical solution of the full system (\ref{U}) and (\ref{V}) has
demonstrated that an arbitrary quiescent ES {\it cannot} be directly
continued into a moving one. Nevertheless, moving ESs exist, but they turn
out to be of {\em codimension two,} i.e., they are {\it double}-isolated,
both in the energy and in the momentum (but, nevertheless, they remain
structurally stable objects). In other words, a moving ES is described by
curves $E(D)$ and $P(D)$. Equivalently, such curves may be represented in
the $(D,c,\chi )$-space, an important characteristic of a moving soliton
being its velocity $c$. The mathematical reason for the codimension 
being two is
that, for the 8th-order model, there are {\em two} pairs of eigenvalues on
the imaginary axis (rather than one pair for the reduced 4th-order model
satisfied by the quiescent ESs). A simple count of dimensions of the
unstable manifold and symmetric set of the reversibility then yields that to
force their intersection (implying the existence of a solitary wave)
requires two parameters to be varied.

The results were found to be sensitive to the value of $\sigma $ (see Eqs.
(1) and (2)); note that in the case $c=0$, $\sigma $ is trivially scaled out 
\cite{we}. The case at which it was easiest to find moving ESs was $\sigma =0
$. The results obtained for this case are summarized in Fig. 1, which shows
that each branch of quiescent ES solutions gives rise, through a pitchfork
bifurcation occurring at some special value of $D$, to two mutually
symmetric branches of moving ESs. In Fig.\ 1 (and Fig.\ 2 below), we cut
each branch at points where they go over into regular (non-ES) solitons (at 
$\chi =-1$). Also, we have not depicted the quiescent branches all the way up
to $\chi =+1$ due to numerical difficulties occurring in this singular
limit. 

In the case $\sigma =0$, it was easy to find additional branches of
moving-ES solutions that are {\em not} connected to the quiescent ones. Only
one such disjoint branch is shown in Fig. 1. It is quite interesting that
this disjoint branch persists for all $|\chi |<1$ without ever bifurcating
from a quiescent ES.

Although the case $\sigma =0$ exactly corresponds to the Thirring model \cite
{Thirring}, it has no straightforward meaning for optical systems.
Therefore, we now focus attention on the most physically relevant case 
$\sigma =1/2$. In this case, only one branch of quiescent ESs, corresponding
to the smallest values of $D$, gives rise, through a bifurcation, to
branches of moving solitons. Scanning the parameter space has not yielded
any disjoint branch, cf.\ Fig. 1. This case is shown, in various forms, in
Fig.\ 2. It is interesting, in particular, that the momentum of the moving
ESs vanishes at a nonzero value of the velocity, exactly (within the
accuracy of the numerical calculation) as it passes into the non-embedded
region ($\chi <-1$), see Fig.\ 2b.

The plot that simultaneously shows the energy of the moving ESs and of the
coexisting quiescent ESs (Fig. 2c) is especially important. Following the
lines of the stability analysis of ESs developed in \cite{YKM}, we can draw
conclusions concerning the stability of both types of the ES solitons. The
analysis developed in \cite{YKM} shows that a small perturbation which {\it
decreases} the energy of an isolated ES solution would trigger a continual
decrease of energy via emission of radiation. In the model considered in
Ref.\ \cite{YKM}, this would eventually lead to complete decay of ESs into
radiation. However, in the present case, a {\em moving} ES is likely to shed
not only its energy, but also momentum, and eventually to decay into a
quiescent ES. Because this instability is weak (sub-exponential), we may
view the full set of ESs as a {\it tri-stable} system, in which transitions
from ESs moving at the velocities $\pm c$ to the quiescent one are possible.

The latter configuration has a potential for use in optical-memory devices.
If an incoming moving ES represents a new bit of information, its
radiation-mediated transition into a quiescent ES can be triggered by a
specially inserted perturbation (e.g.,\ a localized spatial inhomogeneity,
which can be readily made switchable and movable if created by a laser beam
focused on a spot in the medium \cite{Wang}). Thus, the incoming bit could
be captured and stored in the memory.

Further numerical explorations have revealed that the single branch of
moving ESs existing at $\sigma =1/2$ is {\em not} a continuation in $\sigma $
of any branch existing at $\sigma =0$; actually, the continuations of all
those branches terminate between $\sigma =0.1$ and $\sigma =0.2$, but a new
branch appears in the same region  which continues to that found at $\sigma
=1/2$. Continuation of this branch to larger values of $\sigma $ (the case 
$\sigma \rightarrow \infty $ has a physical application to dual-core optical
fibers or waveguides) shows that it terminates at $\sigma \approx 1.645$.
Additional moving ESs exist at still larger values of $\sigma $ (e.g., at 
$\sigma =8.7$), but none was found for $\sigma >10$.

Finally, one can estimate the values of the physical quantities for direct
experimental observation of these ESs in a Bragg-grating medium. First of
all, it is relevant to note that, as Fig. 2a clearly shows, the velocity at
which moving ESs may be observed includes all the values from $0$ up to 
$\sim (1/10)c_{0}$, which is an interesting result by itself, and is quite
convenient for the experiment. 

A parameter which is crucial for the physical relevance of the model
characterizes the relative smallness of the wave (second-derivative) terms
in Eqs.\ (\ref{U}) and (\ref{V}). Obviously, it is $D/W$, $W$ being the ES
width. From the data presented in the insets to Figs.\ 1 and 2, it follows
that this parameter takes a nearly constant value, $\sim 0.1$, along a
moving-ES branch. On the other hand, from the underlying PDEs, it follows
that, in terms of physical quantities, the same smallness parameter is $\sim
\lambda /4\pi c_{0}T$, where $\lambda \equiv 2\pi /k$ is wavelength of
light, and $T$ is the temporal width of the pulse. Taking $\lambda \sim 1.5$ 
$\mu $m, and equating the two expressions for the same smallness, we
conclude that one needs $T\sim 10$ fs.

In recently reported experiments in which the temporal solitons were first
observed in a Bragg-grating medium $T$ was much larger; $\sim 10$ ps \cite
{Krug}. However, much shorter pulses can be produced by means of existing
experimental techniques. For instance, the first experimental observation of
temporal solitons in second-harmonic-generating media used pulses of width 
$58$ ps \cite{temporal}. Moreover, generation of stable pulses with the
temporal duration $_{\sim }^{<}\,5$ fs, which contain just two optical
cycles, has been successfully demonstrated in recent years (see, e.g., Ref. 
\cite{Kalosha} and references therein). This circumstance suggests a
possible link between ESs and rapidly developing studies of the ultrashort
few-cycle optical pulses.

We appreciate valuable discussions with M.J. Friedman, D.J. Kaup, Y.S.
Kivshar, and J. Yang. The stay of B.A.M. at the University of Bristol was
supported by a Benjamin Meaker visiting professorship.

\newpage

\subsection*{Figure Captions}

\begin{figure}[thp]
\caption{ The solution branches of the quiescent and moving (dotted amd
solid lines) embedded solitons in the case $\protect\sigma=0$; squares show
bifurcations, circles label points. (a) The velocity vs.\ the effective
dispersion coefficient $D$. The insets show typical examples of the
solutions on the first branch. (b)
The effective frequency $\protect\chi$ vs.\ $D$.}
\end{figure}

\begin{figure}[thp]
\caption{ Various representations of the single branch of the moving
embedded solitons existing in the physically relevant case, $\protect\sigma 
= 1/2$: (a) the same as in Fig. 1; (b) the momentum vs. $D$; (c) the energy
vs. $D$, with insets showing the solutions at labeled points. The quiescent
solitons branch from which the moving-soliton branch bifurcates is also
shown.}
\end{figure}
\newpage
\setcounter{figure}{0}

\begin{figure}[thb]
\epsfxsize 5in \centerline{\epsffile{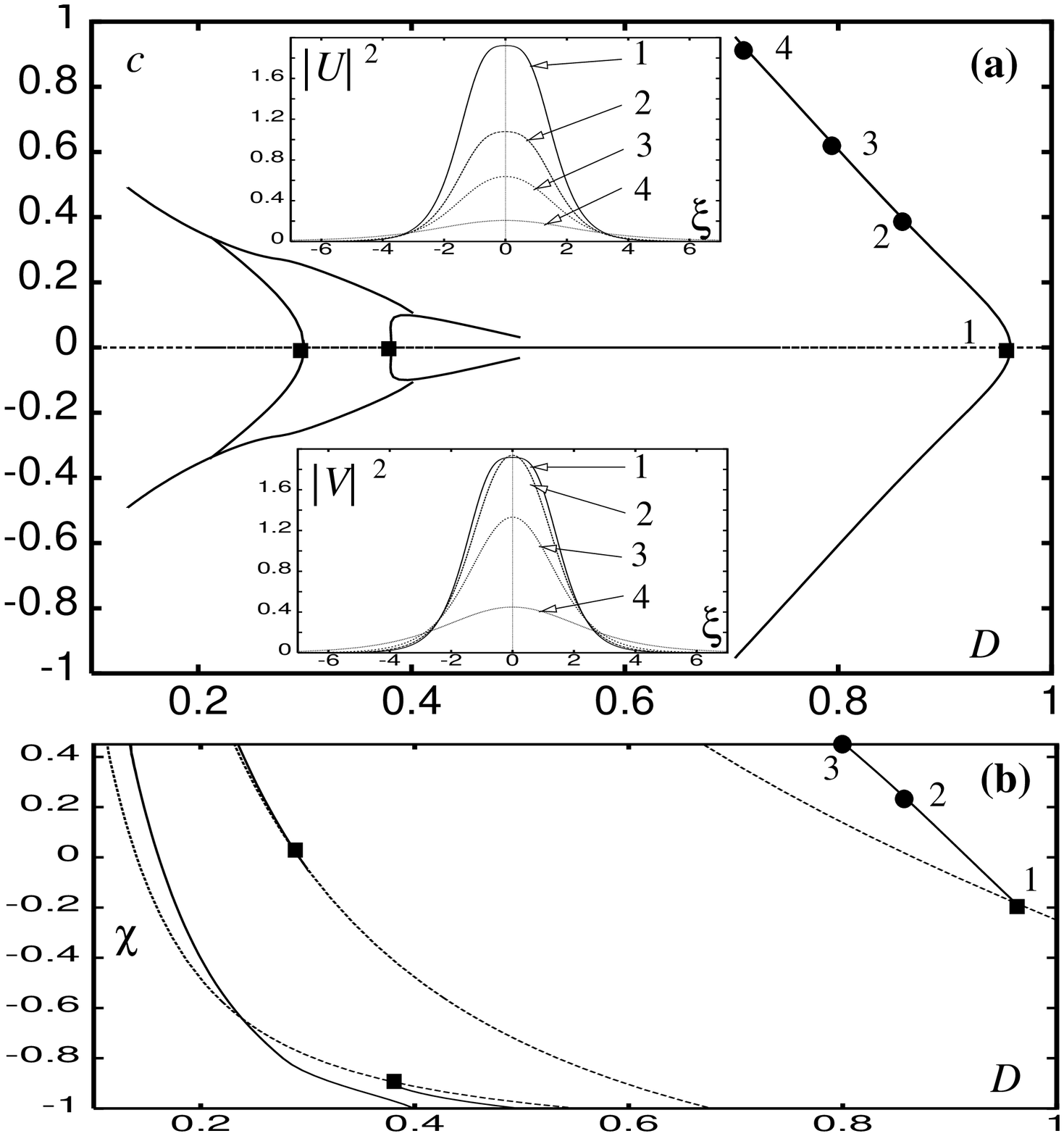}}
\caption{ } 
\end{figure}

\begin{figure}[thb]
\epsfxsize 5in \centerline{\epsffile{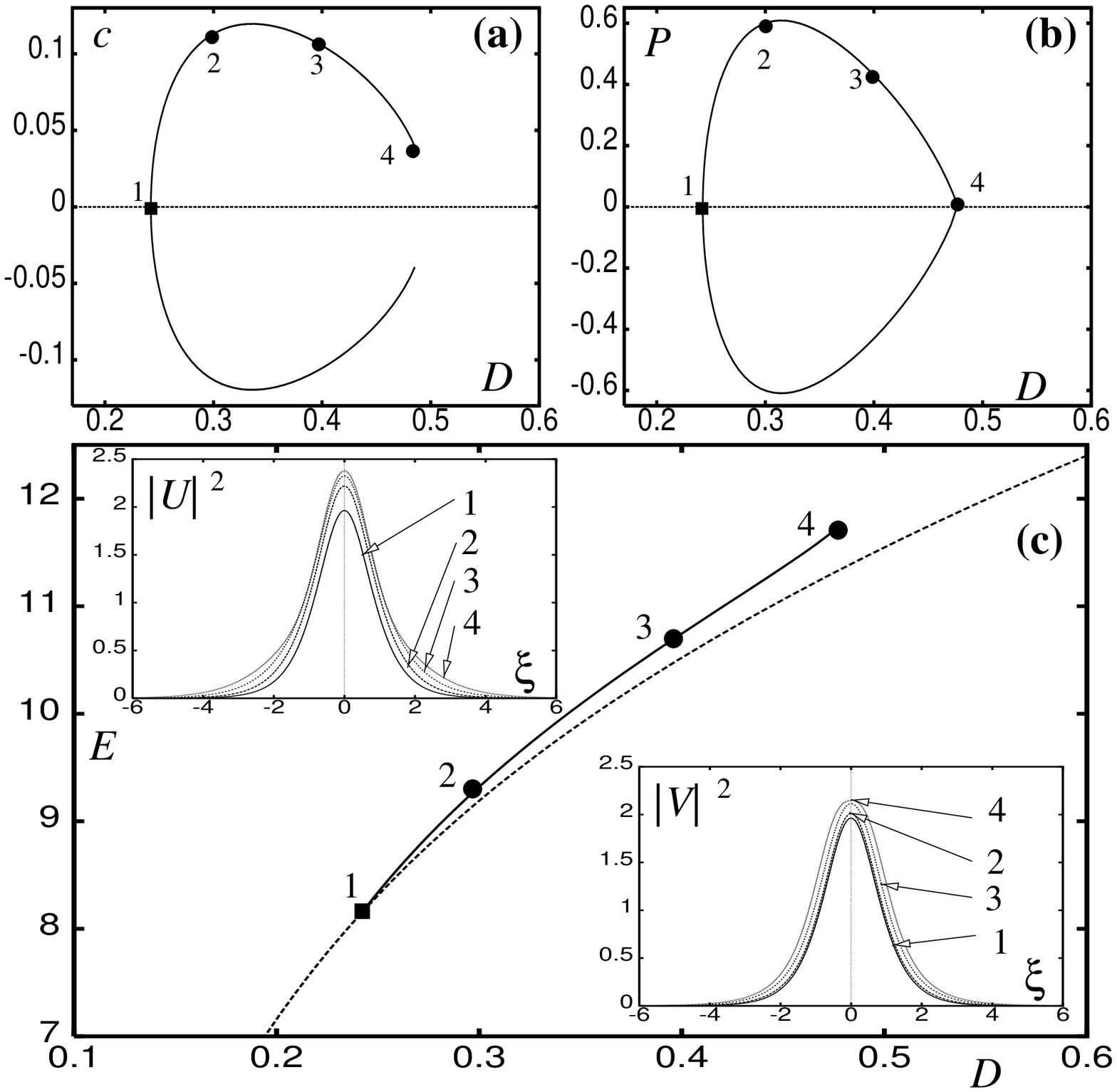}}
\caption{ } 
\end{figure}

\end{document}